\documentclass[pre,aps,epsfig,twocolumn,showpacs]{revtex4}
\usepackage{graphicx}
\begin{document}

\title{The double domain structure of pair contact process with diffusion}

\author{Sungchul Kwon and Yup Kim}
\affiliation{Department of Physics and Research Institute of Basic
Sciences, Kyung Hee University, Seoul 130-701, Korea}

\date{\today}

\begin{abstract}
We investigate the domain structure of pair contact process with
diffusion (PCPD). PCPD is a stochastic reaction-diffusion model
which evolves by the competition of two binary reactions, $2A \to
3A$ and $2A \to 0$. In addition, each particle diffuses
isotropically, which leads to the bidirectional coupling between
solitary particles and pairs. The coupling from pairs to solitary
particles is linear, while the opposite coupling is quadratic. The
spreading domain formed from localized activities in vacuum consists
of two regions, the coupled region of size $R_p$ where pairs and
solitary particles coexist and the uncoupled region of size $R_U$
where only solitary particles exist respectively. As the size of the
whole domain $R$ is given as $R=R_p + R_U$, $R_p$ and $R_U$ are the
basic length scales of PCPD. At criticality, $R_p$ and $R_U$ scale
as $R_p \sim t^{1/Z_p}$ and $R_U \sim t^{1/Z_U}$ with $Z_U > Z_p$.
We estimate $Z_p =1.61(1)$ and $Z_U =1.768(8)$. Hence, the
correction to the scaling of $R$, $Q=R_U /R_p$ extremely slowly
decays, which makes it practically impossible to identify the
asymptotic scaling behavior of $R$. In addition to the generic
feature of the bidirectional coupling, the double domain structure
is another reason for the extremely slow approach to the asymptotic
scaling regime of PCPD.

\end{abstract}

\maketitle

\section{Introduction}
The concept of universality makes possible to understand and
classify the various and complicate critical behavior of a number of
equilibrium models \cite{equi}. For nonequilibrium critical
phenomena, extensive studies during past decades revealed that
various nonequilibrium systems also exhibit universal behavior
characterized by several features \cite{review}. Hence, it has been
an important issue to identify nonequilibrim universality classes by
finding common physical features.

Among nonequilibrium critical phenomena, absorbing phase
transitions (APT's) from fluctuating active states into absorbing
states in which the system is trapped forever have been a field of
growing interest during last decades \cite{review,hin}. Recent
theoretical and numerical studies show that APT's exhibit
universality and it can be classified according to conservation
laws, dimensionality of systems and symmetries of absorbing states
\cite{review,hin,odor}. However only a few universality classes
have been identified so far. Directed percolation (DP)
\cite{hin,odor,Grass,dp-exp} and parity conserving (PC)
\cite{baw4,baw2,nbaw,chate,di,pc2,voter} class are well studied
classes among others. DP class includes systems with no special
attributes except the time reversal symmetry, so that most systems
studied so far belong to this class.

As a research direction to search for further unknown universality
classes, coupled systems have been studied recently
\cite{nbaw,Janssen,nbaw2,nbaw1,dp-arw,noh,alon,png,fungal,uni1,uni2,pcpc,uni3,uni4}.
A coupled system is a multi-species system in which each species is
coupled to the others in certain ways such as bidirectional and
unidirectional coupling in linear or quadratic ways. However the
coupled systems do not always exhibit new critical behavior. For
bidirectionally coupled systems, the critical behavior depends on
the manner of the coupling. For instance, quadratically coupled DP
systems still belong to DP class despite their complex behavior
\cite{Janssen}. However, linearly coupled systems belonging to DP or
PC class exhibit mean-field or non-trivial critical behavior
\cite{nbaw,nbaw2,nbaw1,dp-arw,noh}. Linearly and unidirectionally
coupled systems exhibit new critical behavior at multicritical point
where all sub-systems are critical
\cite{alon,png,fungal,uni1,uni2,pcpc,uni3,uni4}.

Among single species systems, pair contact process with diffusion
(PCPD) can be regarded as a two species system. PCPD has been
extensively studied during last years due to its nontrivial critical
behavior (see \cite{pcpd} for review). However, in spite of
extensive theoretical and numerical studies, the critical behavior
is not clearly uncovered yet. PCPD is a stochastic
reaction-diffusion model, which evolves by the competition of two
processes, fission ($2A \to 3A$) and annihilation ($2A \to
\emptyset$). In addition, each particle performs isotropic
diffusion. Without diffusion, the model is so-called pair contact
process (PCP) belonging to DP class \cite{pcp}. Since the reactions
involve pairs, diffusing solitary particles are not engaged in the
binary reactions. However when two solitary particles form a pair,
the reactions take place. On the other hand, solitary particles are
created from pairs by diffusion. Hence, PCPD can be regarded as a
bidirectionally coupled two species system in which the order of the
coupling is linear in the direction from pairs to solitary particles
and quadratic in the opposite direction. This observation leads to
the cyclically coupled DP and pair annihilation which exhibits the
similar type of critical behavior to that of PCPD \cite{af}.

\begin{figure}
\includegraphics[scale=0.5]{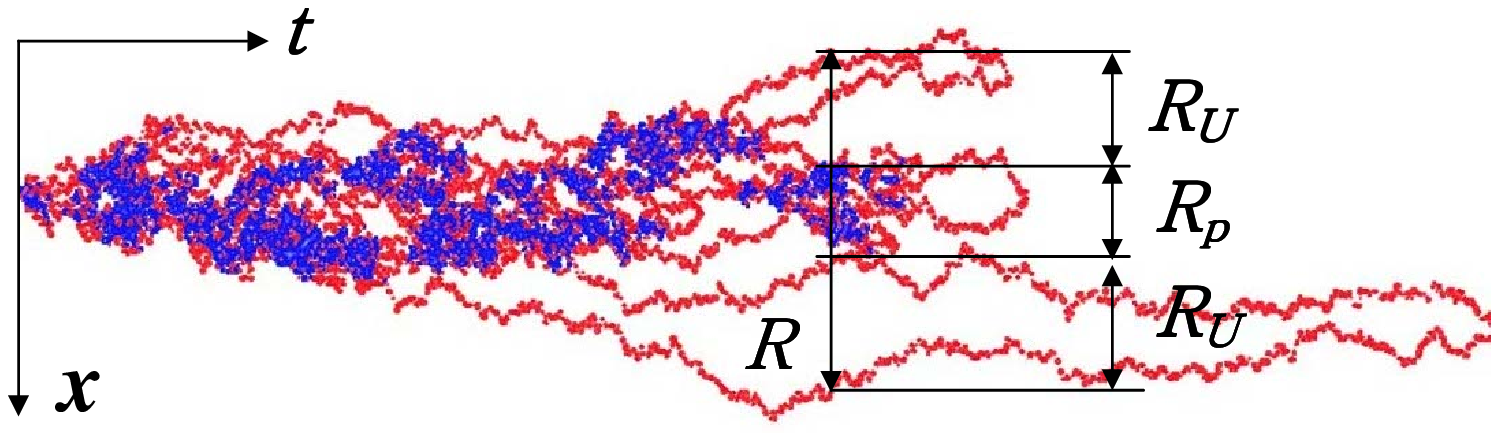}
\caption{\label{pattern}(Color online) The double domain structure
of PCPD. The region with size $R_p$ is the coupled region in which
pairs (blue) and solitary particles (red) coexist. The region with
size $R_U$ is the uncoupled region in which only solitary particles
exist.
 }
\end{figure}

In this paper, we investigate the domain structure of PCPD. When a
spreading domain is formed from localized initial activities, the
quadratic coupling from solitary particles to pairs allows the
pair-free region in which only solitary particles are present. We
call the pair-free region so-called {\it uncoupled} region. On the
other hand, the linear coupling from pairs to solitary particles
results in the so-called {\it coupled} region in which pairs and
solitary particles coexist. The pair-free region encloses the
coupled region as shown in Fig.~\ref{pattern}. Hence, the spreading
domain is divided into two regions, coupled and uncoupled region.
This kind of double domain structure was found in unidirectionally
coupled two level hierarchies \cite{uni3,uni4}.
As shown in previous studies on unidirectionally coupled systems
\cite{uni1,uni2,pcpc,uni3,uni4}, the measurements of critical
exponents are very difficult due to long-time drift of the
exponents. The one reason of the drift is the generic feature of the
unidirectional coupling \cite{uni1,uni2}, the other is the double
domain structure \cite{uni3,uni4}. In measuring critical exponents,
one can overcome the latter effect by measuring quantities in each
region separately.

To explain the effect of the domain structure on the critical
behavior, let us consider the unidirectionally coupled DP-PC
coupling \cite{uni4}. In the DP-PC coupling, PC process ($B$) is
unidirectionally coupled with DP process ($A$) via $A \to A+2B$
(linear). Since DP process spreads faster than PC process, the
coupling changes the spreading behavior of PC process. There are no
interactions between $A$ and $B$ species. Starting with a single $A$
particle on $A$-level, a $B$-domain is form and spreads faster than
a $A$-domain does due to the coupling. As a result, the uncoupled
region where no $A$ particles exist is formed \cite{uni3}. The
coupled region is just the region of $B$-domain overlapped with the
$A$-domain where the coupling exists. Hence, the size of the
$B$-domain ($R_B$) is the sum of the size of coupled ($R_C$) and
uncoupled region ($R_U$), $R_B = R_C + R_U$. At the multicritical
point where both species are critical, the size of each domain
increases in power-law such as $R_A \sim t^{z_A}$, $R_B \sim
t^{z_B}$, $R_C \sim t^{z_C}$ and $R_U \sim t^{z_U}$. Since $z_C =
z_A$ by definition, the scaling of $R_B$ is given as $R_B \sim a
t^{z_A} +b t^{z_U}$, $a$ and $b$ are constants. A recent study
showed $z_A \geq z_U$ for the fast spreading $A$-domain (source)
\cite{uni4}. Hence, $R_B$ scales as $t^{z_A}$ asymptotically for
faster spreading source $A$, and $R_U$ plays the correction to the
scaling of $R_B$. Since the difference $z_A - z_U$ is sufficiently
small in the DP-PC coupling, the existence of the uncoupled region
makes precise estimates of $z_B$ difficult within moderate
simulation time. For the DP-PC coupling, the estimate of $2z_B$ is
$1.24(1)$ which is less than the expectation $2z_B = 2z_{DP} =1.265$
\cite{uni4}. On the other hand, the estimate of $2z_U$ is $2z_U =
1.20(2)$ which is compatible with the leading scaling.

As in unidirectionally coupled systems, it is expected that the
double domain structure of PCPD also makes it difficult to identify
the critical spreading behavior precisely. The aim of this paper is
to investigate the effect of the double domain structure on the
critical spreading of PCPD and to estimate critical exponents more
precisely.
We consider the PCPD of Ref. \cite{hhpcpd}. In this model, a
randomly selected particle attempts to hop to one of the nearest
neighbor sites with an equal probability. If the target site is
empty, the attempt is accepted. However if the target site is
occupied, (i) two particles annihilate with probability $p$ or (ii)
the hopping attempt is rejected and the pair (the chosen particle
and one at the target site) tries to create a particle at the
randomly chosen nearest neighbor site of the pair. When the target
site is occupied, the branching attempt is rejected. The critical
point of this model is $p_c = 0.133\;519(3)$ in one dimension
\cite{hhpcpd}.

PCPD has three sectors in configuration spaces according to the
existence of pairs($P$) and solitary particles($S$). The one is the
configurations in which both pairs and solitary particles are
present ($PS$-ensemble). We call configurations with at least one
pair (two solitary particles) $P$-ensemble ($S$-ensemble). In
$P$-ensemble ($S$-ensemble), solitary particles (pairs) may be
present or not. The $P$-ensemble is the reactive subspace of Ref.
\cite{dickman}. A conventional ensemble includes configurations with
at least two particles which can be either two solitary particles or
one pair. We call the conventional ensemble $All$-ensemble. Since
solitary particles are effectively linearly coupled to pairs, the
existence of pairs implies the existence of solitary particles.
Hence, $PS$-ensemble should coincide with $P$-ensemble
asymptotically. However, solitary particles transform into pairs by
collisions, the coupling in this direction is quadratic. Hence, the
existence of solitary particles does not always guarantee the
presence of pairs due to the long life time of solitary particles.
So $S$-ensemble coincide with $All$-ensemble. As a result, there are
two distinct ensembles in PCPD.
\begin{figure}
\includegraphics[scale=0.4]{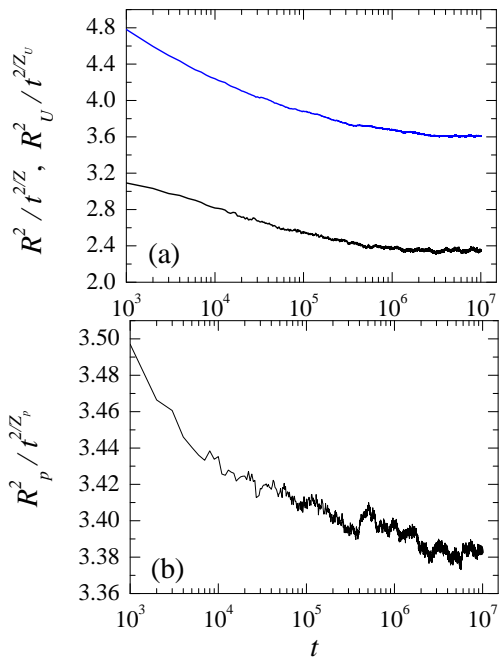}
\caption{\label{range}(Color online) The scaling plots of various
sizes. (a) The scaling plot of $R^2$ (black) and $R_{U}^2$ (blue)
with $Z=1.663$ and $Z_U = 1.768$. (b) the scaling plot of $R_{p}^2$
with $Z_p = 1.61$. }
\end{figure}

We define the size of each domain as follows (See
Fig.~\ref{pattern}). The size of the whole domain at time $t$
($R(t)$) is defined as the distance between the leftmost and the
rightmost particle. When both pairs and solitary particles exist
simultaneously, we can define the size of the coupled region ($R_C
(t)$) and the size of the uncoupled region ($R_U (t)$). Since
solitary particles are linearly coupled to pairs, we define the
size of the coupled region ($R_C (t)$) as the spreading distance
of pairs ($R_p$) defined as the distance between the leftmost and
the rightmost pair. Then, $R_U (t)$ is given as $R_U = R - R_p $.
Hence, we have three different lengths, $R$, $R_p$ and $R_U$ in
PCPD. To take into account the three length scales at the same
time, one should use $PS$-ensemble in which only two lengths,
$R_p$ and $R_U$ are the fundamental length scales of PCPD due to
$R=R_p + R_U$.

\begin{figure}
\includegraphics[scale=0.4]{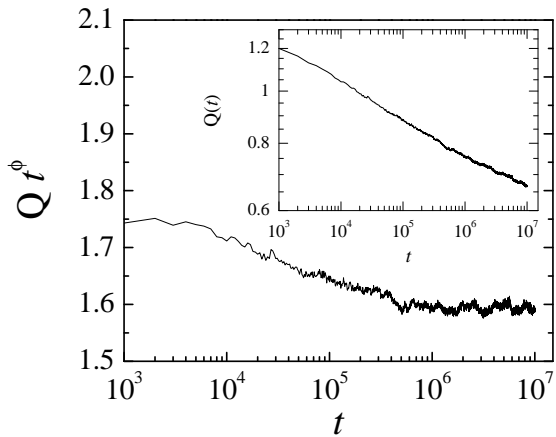}
\caption{\label{ratio} The scaling plot of the ratio
$Q=R_{U}/R_{p}$. The main plot shows $Q t^{\phi}$ with $\phi =
0.054$. The inset shows the double logarithmic plot of $Q(t)$. }
\end{figure}

At $p_c = 0.133\;519$, we perform defect Monte Carlo simulations
with a pair on one dimensional empty lattice. We run simulations up
to $t=10^7$ time steps using $3.6\times 10^6$ independent runs. We
measure the squared sizes, $R^2$, $R^{2}_p$ and $R^{2}_U$ for
surviving $PS$-ensemble. The PCPD has two absorbing states, vacuum
and states with one diffusing solitary particle. Hence, we stop the
simulations when the total number of particles $N$ is less then two.
At the criticality, the squared sizes scale as $R^2 \sim t^{2/Z}$,
$R^{2}_{p} \sim t^{2/Z_p}$ and $R^{2}_{U} \sim t^{2/Z_U}$.
Fig.~\ref{range} shows the scaling plots of the squared sizes,
$R^{2}_x / t^{2/Z_x }$. We obtain the best scaling plot with
$Z=1.663(5)$, $Z_p = 1.61(1)$ and $Z_U = 1.768(8)$ respectively. The
errors of our estimates should be larger due to the error of $p_c$.
Within the numerical errors at the criticality, our estimate of $Z$
agree with the previous studies \cite{pcpd}, especially $Z=1.70(5)$
\cite{chate}. Since $Z_U > Z_p$, the total spreading distance
$R(=R_{p}+R_U )$ should scale as $R \sim t^{1/Z_p}$, and $R_U$ plays
the role of the correction to the scaling as in unidirectionally
coupled systems. Hence, we conclude $Z=Z_p = 1.61(1)$ which is the
smallest value among the estimates of previous studies \cite{pcpd}.
For reference, we also measure $R^2$ using $All$-ensemble. The
difference of $All$-ensemble from $PS$-ensemble or $P$-ensemble is
that $All$-ensemble includes configurations without pairs. Since
solitary particles spread diffusively, it is expected that $R$
averaged over $All$-ensemble scales differently from that of
$PS$-ensemble. From the scaling plot of $R^2 /t^{2/Z}$, we estimate
$Z=1.676(3)$ for $All$-ensemble which is larger than that of
$PS$-ensemble (not shown). The slow spreading of the whole domain in
$All$-ensemble results from the diffusive motions of solitary
particles in configuration without pairs.

Since $R$ scales as $R=R_{p}(1+R_U /R_p )$, the correction to the
$R$ is $Q=R_{U}/R_{p}$ which decays as $Q \sim t^{-\phi}$ with $\phi
= (Z_U - Z_p )/Z_p Z_U$. To see how the correction decays slowly in
time, we plot $Q$ in Fig.~\ref{ratio}. When $Q \ll 1$, the
correction is negligible. However, as shown in the inset, $Q$ is
still comparable to one even at $t=10^7$. We obtain the best scaling
plot of $Q t^{\phi}$ with $\phi = 0.054(4)$. As $Q$ decays with very
small $\phi$, it is practically impossible to reach the asymptotic
scaling regime of $R \sim R_p$. As a result, one should take the
double domain structure into account for the more precise
measurement of the dynamic exponent $Z$.

\begin{figure}
\includegraphics[scale=0.4]{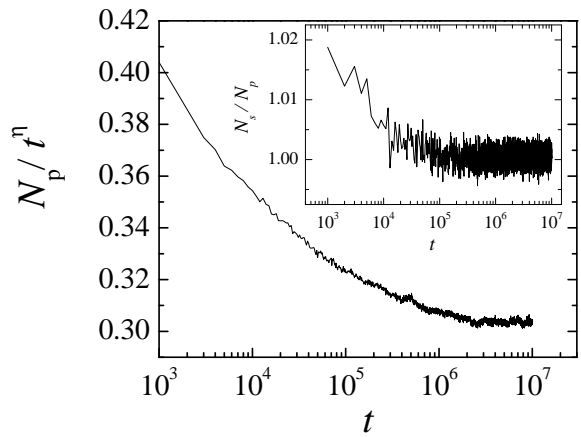}
\caption{\label{eta} The scaling plot of $N_p$. The main plot shows
$N_p /t^\eta$ with $\eta=0.275$. The inset shows the semilogarithmic
plot of the ratio $N_s /N_p$.
 }
\end{figure}

\begin{figure}
\includegraphics[scale=0.4]{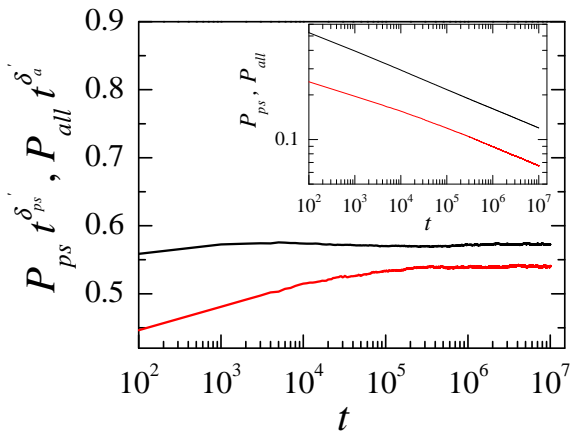}
\caption{\label{p}(Color online) The scaling plot of $P_{ps}$ and
$P_{all}$. The main plot shows the scaling plot of $P_{ps}
t^{\delta^{\prime}_{ps}}$ (red) and $P_{all}
t^{\delta^{\prime}_{a}}$ (black) with $\delta^{\prime}_{ps} =
\delta^{\prime}_{a}=0.13$. We add a constant $-0.4$ to the scaling
plot of $P_{all}$ for the better presentation. The inset shows the
double logarithmic plots.
 }
\end{figure}

In addition to the sizes, we also measure the number of pairs
($N_p$) and of solitary particles ($N_s$) averaged over all samples,
and the survival probability of $PS$-ensemble ($P_{ps}$) and
$All$-ensemble ($P_{all}$). As solitary particles are linearly
coupled to pairs, $N_s$ is proportional to $N_p$. At criticality,
$N_p$ scales as $N_p \sim t^{\eta}$. Fig.~\ref{eta} shows the
scaling plot $N_p /t^\eta$. We obtain the best scaling plot with
$\eta = 0.275(5)$. The inset show the ratio of $N_s /N_p$, which
converges to one as expected. $P_{ps}(t)$ and $P_{all}(t)$ decay in
power-law as $P_{ps} \sim t^{-\delta^{\prime}_{ps}}$ and $P_{all}
\sim t^{-\delta^{\prime}_{a}}$ at the criticality. Fig.~\ref{p}
shows the scaling plots of $P_x t^{\delta^{\prime}_x}$ and the
double logarithmic plots of $P_{ps}$ and $P_{all}$. With
$\delta^{\prime}_{ps}=\delta^{\prime}_{a}=0.130(3)$, we obtain the
best scaling plots. The system escapes from the non-reactive sector
in which only diffusing solitary particles are present via pair
annihilations of solitary particles. As an naive argument for the
equality of $\delta^{\prime}_{ps}=\delta^{\prime}_{a}$, the escaping
probability from non-reactive sector within time $\tau$ may scale as
$1-\tau^{-1/2}$ which is the death probability of two diffusing
particles undergoing the reaction $S+S \to 0$ within $\tau$. Hence,
the contribution of non-reactive sector to $P_{all}$ is negligible
due to the fast escaping probability, which leads to the same
scaling behavior of $P_{all}$ as $P_{p}$.

In summary, we investigate the domain structure of PCPD. We
numerically confirm the existence of the double domain structure
in PCPD. This double domain structure comes from the linear and
quadratic bidirectional couplings. The structure intrinsically
makes the serious correction to the scaling of the critical
spreading of a domain.

Starting with a pair, a domain grows and spreads in vacuum via
fission ($A+A \to 3A$) and spontaneous annihilation of pairs ($2A
\to 0$) in PCPD. In addition to the binary reactions, each particle
diffuses isotropically, which leads to the bidirectional coupling
between solitary particles and pairs. The coupling from pairs to
solitary particles is linear, while the opposite coupling is
quadratic. The difference of the coupling ways results in the double
domain structure of the whole domain, the coupled and the uncoupled
region respectively. As a result, the size of the whole domain ($R$)
is given as the sum of the size of the coupled region ($R_p$) and of
the uncoupled region ($R_U$). Hence, $R_p$ and $R_U$ are the basic
length scales characterizing the scaling behavior of the spreading
domain in PCPD. We numerically find that $R_p$ and $R_U$ scale as
$R_p \sim t^{1/Z_p}$ and $R_U \sim t^{1/Z_U}$ with $Z_U > Z_C$ at
criticality. Hence, $R$ should asymptotically scale as $R \sim
t^{1/Z}$ with $Z=Z_p$ and $R_U$ plays the role of the correction to
the scaling. However, the direct measurement of $R$ leads to the
underestimate of the asymptotic value of $Z$ because the correction
$Q=R_U /R_p$ decays with very small exponent. Since it is
practically impossible to reach the asymptotic scaling region of $R
\sim t^{1/Z_p}$, it is important to take the domain structure into
account in simulations for more precise estimate of the dynamic
exponent $Z$ of PCPD.

We classify particle configurations into four ensembles, which are
finally reduced to two distinct ensembles, $P$-ensemble and
$All$-ensemble respectively. $All$-ensemble includes configurations
without pairs, while $P$-ensemble does not. The survival
probabilities of two ensembles decay with the same exponent.
However, the whole domain appears to spread more slowly in
$All$-ensemble than in $P$-ensemble due to the diffusive motions of
solitary particles in configurations without pairs. Hence, in
addition to the domain structure, the diffusive motions of solitary
particles in $All$-ensemble raise another correction to the scaling
of the total spreading distance $R$ which does not appear in
$P$-ensemble.

As the linear-quadratic bidirectional coupling is the common
feature of various PCPD studied so far, the double domain
structure should appear in other PCPD variants. Among PCPD
variants, we investigate the domain structure of the bosonic PCPD
with soft-constraint of Ref. \cite{chate} which is known to
exhibit the clear power-law decays at the criticality. For this
model, we also confirm the existence of the double domain
structure and the critical spreading behavior similar to that of
PCPD studied in this paper. Hence, the double domain structure is
a common feature of PCPD variants.

As shown in recent studies, as PCPD extremely slowly approaches to
its asymptotic scaling region, it is very difficult to identify the
critical behavior. The bidirectional coupling should be the main
reason in itself. On the other hand, the double domain structure
naturally appears due to the linear-quadratic bidirectional
couplings, which enhance the long-time drift of dynamic exponent
$Z$. One can overcomes the latter effect by considering the scaling
behavior of sub-domains separately as in unidirectionally coupled
systems.

\begin{acknowledgements}
This work was supported by Grant No. R01-2004-000-10148-0 from the
Basic Research Program of KOSEF.
\end{acknowledgements}

\end{document}